\documentclass[aps,pre,twocolumn,groupedaddress,showpacs]{revtex4}
\newif\ifpdf            
\ifx\pdfoutput\undefined \pdffalse \else \pdftrue \fi
\ifpdf 
\usepackage[pdftex]{graphicx}
\pdfcompresslevel9 \DeclareGraphicsExtensions{.jpg,.pdf,.png,.mps}
\else 
\usepackage[dvips]{graphicx}
\DeclareGraphicsExtensions{.eps,.ps}
\fi
\usepackage{bm}

\begin{document}

\title{Chirality transfer and stereo-selectivity of imprinted cholesteric networks}

\author{S.~Courty, A.R.~Tajbakhsh and E.M.~Terentjev}

\affiliation{Cavendish Laboratory, University of Cambridge,
Madingley Road, Cambridge, CB3 0HE, U.K. }

\date{\today}

\begin{abstract}
Imprinting of cholesteric textures in a polymer network is a
method of preserving a macroscopically chiral phase in a
system with no molecular chirality. By modifying the elastics properties
of the network, the resulting stored helical twist can be
manipulated within a wide range since the imprinting efficiency
depends on the balance between the elastics constants and twisting
power at network formation. One spectacular property of
phase chirality imprinting is the created ability of the network to
adsorb preferentially one stereo-component from a racemic mixture. In
this paper we explore this property of chirality transfer from a macroscopic
to the molecular scale. In particular, we focus on the
competition between the phase chirality and the local
nematic order. We demonstrate that
it is possible to control the subsequent release of chiral solvent
component from the imprinting network and the reversibility of the
stereo-selective swelling by racemic solvents.
\end{abstract}


\maketitle

\section{Introduction}

Since its first discovery in the middle of the $19$ century by
Pasteur \cite{pasteur} and attempts on mathematical abstraction by
Kelvin \cite{kelvin}, chirality fascinates the scientific
community across the disciplines. The nature appears to be
inherently chiral. From the atomic scale with asymmetric carbon,
to much larger length scales -- like our hands and up to spiral
galaxies, all have the same common feature of lacking the
inversion symmetry, while not characterized by any vector
(dipolar) property. In other words, many natural objects are
non-superimposable with their mirror image and define a pair of
opposite handedness,  right and left. It is important to realize
that handedness is not an absolute concept; its quantitative
characteristics depend on the property being observed
\cite{Osipov95,Harris99}, the origin of many questions and
disagreements between different groups of results.

On a fundamental level, chirality is at the origin of a the
``paradox of life''. In chemistry, molecules are equally present
in their different forms, right and left-handed (this distinction,
arbitrary in general, could be based on the rotation of plane
polarization of a beam induced by the particular molecule under
consideration). However, living system use only one stereo-form:
amino acids constituting proteins are all left-handed and all
nucleic acids of DNA and RNA are right-handed: a phenomenon of
homochirality. Recently it has been demonstrated that a
non-polarized light flow coupled with a parallel magnetically
field induce one optical form from a racemic solution (ref). This
magnetochiral anisotropy has revived a debate of the early sixties
about the origin of homochirality.

``Wrong-handed'' enantiomers can have dramatic consequences in all
aspects of life. The left-handed sugar (L-glucose) tastes just as
sweet as the right-handed one (D-glucose), but your body can't use
it as an energy source. Ibuprofen is one example of the chiral
drugs widely used in pain relievers; the left-handed version is
about four times as strong as the right-handed enantiomer. A
famous medical disaster of 1960s involved the sedative
thalidomide, initially produced with no chiral discrimination --
it was later shown that one handedness of it had a poisonous
effect. Today it is crucial for any pharmaceutical product to
control the chirality, which implies the ability to synthesize
selectively one enantiomer, or to separate the left- from the
right-handed form and finally to quantify the stereo-selectivity.
But how to separate a pair of opposite left- and right-handed
molecules which differ only in a subtle way? This question
represents a challenging problem. The main difficulty for
stereo-selection is the weakness of molecular interaction
sensitive to handedness. One of the main techniques in
stereo-separation is column chromatography, in which a racemic
mixture diffuses through a silica gel coated with a molecular
layer of specific chirality: the two enantiomers of the mixture
diffuse at slightly different rates due to the weak van der Waals
attraction to the gel coating (due to the chiral corrections to
high-order dielectric polarizability). New methods have been
developed to measure such forces, for example, by detecting a
difference in adhesion between an AFM tip coated with chiral
molecules and the left- or right-handed substrate
\cite{McKendry98}. In all cases, the methods of stereo-selection
have been based on interactions of individual molecules.

Recently a new approach has been suggested \cite{Mao:01b}, based
on the macroscopic phase chirality in topologically imprinted
cholesteric networks \cite{Courty:03}. Cholesteric order in liquid
crystals results in a breaking of inversion symmetry of the
nematic phase due to the presence of chiral molecules, e.g. a
dopant in a mainly nematic material. As an immediate consequence,
the director $\bm{n}$ in cholesteric phase is spontaneously
twisting in space, in a periodic helical fashion. The macroscopic
helical pitch $p=2\pi/q_{0}$ is inversely proportional to the
concentration $\phi$ of the chiral dopant. The cholesteric phase
frozen in the permanent polymer network (an elastomer or a gel)
has been known for a long time
\cite{Finkelmann:81,Freidzon:86,Zentel:87,Meier:90}. However, the
topological imprinting of phase chirality is a new concept. In
1969, de Gennes was the first to suggest that chiral order can be
preserved by cross-linking a conventional polymer in a chirally
doped liquid crystal phase \cite{deGennes:69a}. Such a network
would retain a memory of the phase chirality even when the dopant
is completely removed, leaving an internally stored helical twist
in a material without any molecular lack of symmetry. One has to
emphasize the difference with (common in biochemistry) approach to
imprint a specific molecular property, such a site for enzyme to
attach -- here the imprinted chiral property is on the macroscopic
scale and affects collective, coherent properties of the system.
Experimentally, elements of chiral imprinting have been
demonstrated in different polymer systems
\cite{Tsutsui:80,Hasson:98}.

How to monitor the current state of phase chirality, the helical
pitch $p$ in the imprinted network? A traditional method used in
studies of liquid cholesterics is based on the selective
reflection of light at a certain wavelength. However, in most
techniques it is not chirality specific (only explores the length
scale matching between the helix and the light), and also requires
very thin samples with high optical quality -- whereas in practice
even the best cholesteric rubbers never have such a quality. In
fact, recent studies has shown that one can generate the bandgap
for both right- and left- circular polarizations of incoming light
by only a slight mechanical deformation of cholesteric elastomers
\cite{Bermel:02,Cicuta:02}. Instead we choose to measure the
optical rotation $\Psi$ (the angle of rotation of plane
polarization of light propagating along the helical pitch), which
is highly sensitive to any small variation of the helical twist
related to any change in chiral molecules concentration $\phi$ in
the network.

One particular property of imprinted network is the
stereo-selection between left- and right-handed molecules from a
racemic mixture which is used as a solvent \cite{Courty:03}. The
imprinted network will preferentially absorb and retain chiral
molecules fitting the handedness of its imprinted macroscopic
helix, in order to restore its initial helical twist $q_{0}$.
However, this stereo-selectivity effect is highly non-trivial and
influenced by several competing factors, in particular by the
local nematic order described by the parameter $Q$. In a typical
thermotropic liquid crystal $Q$ depends on temperature difference
$[T-T_c]$, with the critical point $T_c$ a function of material
composition, in particular, the solvent content. As one adds a
solvent to a liquid crystalline gel, the magnitude of the local
nematic order parameter changes -- usually decreases with the
overall solvent concentration $\phi$ in the network. This results
in a rapid change in local optical birefringence $\Delta m$
(affecting the optical rotation $\Psi$) and also the strength of
phase chirality (reducing the specific interaction with chiral
solvent). As soon as the material becomes isotropic, i.e. loses
its coherent cholesteric structure altogether, it also loses the
stereo-selectivity (at least to the accuracy of our detection
methods). In this paper we will explore the ability of
stereo-selective swelling of topologically imprinted networks by
studying the competition between this local liquid crystalline
order and the macroscopic phase chirality. Finally, we will show
by manipulating the phase chirality (e.g with temperature) it is
possible to control the subsequent release of chiral solvent
component from the imprinting network and the reversibility of the
stereo-selective separation.
\begin{figure}
\resizebox{0.3\textwidth}{!}{\includegraphics{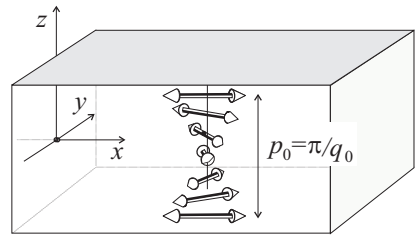}}
\caption{Spatial distribution of the director $\bm{n}$ (shown here
by double-headed arrows) in an ideal cholesteric helix along the
macroscopic optical axis $z$.} \label{helix}
\end{figure}
%

\section{Imprinting of phase chirality}
In theoretical analysis of chiral imprinting \cite{Mao:00}, Mao
and Warner (MW) have introduced a control parameter that measures
the (inverse) strength of imprinted helicity in the polymer
network, $\alpha = \sqrt{K_2/D_1}q_0$, where $K_2$ is the Frank
(twist) elastic constant \cite{deGennes93}, $q_0$ is the helix
wavenumber at network formation (a measure of its twisting power)
and $D_1$ is the relative-rotation coupling constant \cite{WT03}.
If the network is formed with a large $\alpha$, it would not be
able to sustain its helical twisting when the chiral dopant is
removed, while at $\alpha \ll 1$ the more rigid elastic network
retains most of the imprinted helix. In agreement with MW theory,
we found that the stability of topological imprinting of phase
chirality is a function of the chiral order parameter $\alpha$
(which can be altered by modifying the crosslinking density which
is directly related to the elastic constant $D_{1}$).

\begin{figure}
\resizebox{0.32\textwidth}{!}{\includegraphics{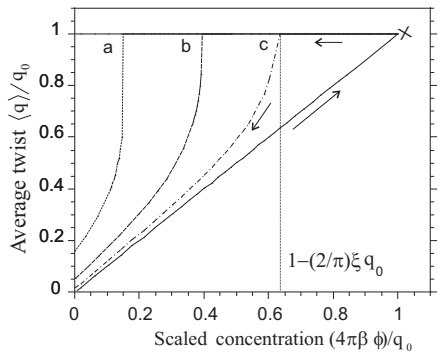}}
\caption{Relative helical wave number as function of current
chiral dopant concentration $\phi$ for the MW model. On increasing
$\phi$ the wave number increases; the crosslinking occurs at
$q=q_0$, a point labelled by the cross. On subsequent decreasing
$\phi$ the strength of imprinting depends on parameter $\xi q_0$:
curves ({\sf a}) $\xi q_0=0.75$, ({\sf b}) $\xi q_0=1.05$, ({\sf
c}) $\xi q_0=1.75$.  } \label{models}
\end{figure}

At scales below the pitch length, cholesterics are an amorphous
uniaxial medium, described by the local nematic order parameter
$Q_{ij}= Q(T,\phi) (n_{i}n_{j}-\frac{1}{3}\delta_{ij})$, with the
director $\bm{n}$ a periodic modulated function of coordinates, in
the ideal state rotating along a single axis $z$:
$n_{x}=\cos\theta, n_{y}=\sin\theta, n_{z}=0$. In the ideal
cholesteric the azimuthal angle is $\theta=q_{0}z$, with the
corresponding helical pitch $p_{0}= \pi/q_{0}$, Fig.~\ref{helix}.
This spontaneously twisted director distribution can be due to the
presence of chiral molecules in the nematic polymer network during
its crosslinking. If, after crosslinking, the chiral dopant is
removed from this network (or replaced by an achiral solvent), two
competing processes occur: The Frank energy penalty for the
director twist, $\frac{1}{2}K_{2} (\bm{n}\cdot curl \bm{n})^{2}$,
demands the cholesteric helix to unwind; any remaining twist
causes the rise of Frank energy. The local anchoring of the
director to the rubbery network, measured by the relative-rotation
coupling constant $D_{1}$, resists any director rotations, thus
acting to preserve the originally imprinted helix. $D_1$ is
proportional to the rubber modulus of the network and, through it,
to the crosslinking density. The free density energy is then given
by the competition of two effects:
\begin{equation}
F = \int {\textstyle{\frac{1}{2}}} \left[ K_{2}
({\textstyle{\frac{d}{dz}}}\theta - q)^{2} +D_{1}\sin^{2} (\theta-
q_{0}z) \right] dz  \label{free energy}
\end{equation}
with $q$ the helical wave number that the current concentration of
chiral solvent $\phi$ would induce, $q=4\pi\beta\phi$, where
$\beta$ is the microscopic twisting power of the solute
\cite{deGennes93}. With its complete removal, $\phi=0$ and $q=0$,
while the concentration at crosslinking is taken as $\phi_0$, with
$q=q_0$, see Fig.~\ref{models}. MW have quantified the balance
between these two opposing trends by introducing a parameter
$\alpha$, which in this case reads:
\begin{equation}
\alpha= \xi [q_{0}-q(\phi)] \, , \ \ \ \ {\rm with} \ \ \xi
=\sqrt{K_{2}/D_{1}}  \label{alpha}
\end{equation}
the nematic rubber penetration length \cite{WT03}. Note that both
$K_2$ and $D_1$ are proportional to the square of local nematic
parameter $Q$, and so the length $\xi\approx \, $const. The
wavenumber $q(\phi)$ in this definition is a linear function of
chiral dopant concentration in the current state. The resulting
classical problem of elliptical functions predicts that the helix
coarsens and its period increases, as soon as $\alpha$ increases
past the threshold value of $\pi/2$.

Fig.~\ref{models} illustrates the model results by plotting the
ratio $\langle q \rangle / q_{0}$, which is the relative number of
remaining cholesteric phase inversions (helix periods). After
crosslinking at $\{\phi = \phi_{0}, q=q_0 \}$, on removing the
chiral dopant the network initially is not affected until a
critical value $4\pi\beta \, \phi^* = q_{0}-2/(\pi\xi)$ is
reached. This critical point may or may not be accessed on
de-swelling, depending on the values of $\xi$ and $q_0$. The value
of $\langle q \rangle$ still remaining at $\phi=0$ is the amount
of topological imprinting of helix by the network. Very low
crosslink density leads to low $D_1$, high $\xi$ and the nearly
complete unwinding of helices (loss of imprinting). A highly
crosslinked network, leads to low $\xi$ and, if $\phi^* \le 0$,
the complete retention of the original helix.

Our purpose in this paper is to explore the stereo-selectivity of
polymer networks with no molecular chirality, but the phase
chirality imprinted in the way described above. Stereo-selectivity
leads to the imbalance $\Delta\phi$ of chiral enantiomers swelling
the network, which we monitor through the weight and shape of the
sample (providing the data on total $\phi$), and the changes in
optical rotation (giving direct access to $\Delta \phi$).

\begin{figure}
\resizebox{0.4\textwidth}{!}{\includegraphics{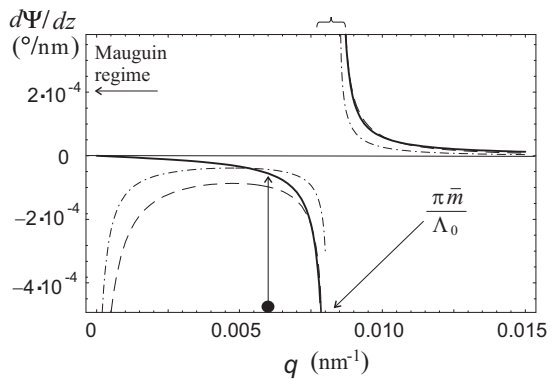}}
\caption{The rate of optical rotation $d\Psi/d z$, as function of
the helix wavenumber $q$. The
solid line shows the interpolated result with correct limiting
behavior, from \cite{Courty:03b}. The broken lines show the
classical de Vries plots for decreasing
local birefringence ($\Delta m =0.15$ and $\Delta m =0.05$).
The bandgap is at $\lambda' =1$, with a width
decreasing with $\Delta m$. The dot marks the initial cholesteric
pitch $p_0 =\pi/q_0 =496$\,nm.} \label{deVries plot}
\end{figure}

In order to interpret the results on optical rotation, we shall
need to analyze two different regimes. Weak optical rotation
(Faraday effect) of a solution with small chiral imbalance is a
simple, unambiguously linear function of $\Delta\phi$. However,
when measuring the changes in the rotation of plane polarization
of light passing through the cholesteric helix (whether imprinted
or natural), a more delicate analysis is required. For this we
need to revise the classical results of de Vries
\cite{deVries:51,Belyakov79} on the rotatory power of a
cholesteric helix.  The details of its application to this
experimental problem are described in greater in our earlier work
on cholesteric elastomers \cite{Courty:03b}. The main issue in a
photonic bandgap system such as the cholesteric helix is the
highly nonlinear rotation rate, whose value, and even sign,
strongly depend on the relation between wavelength of light and
the pitch $p=\pi/q$. The non-dimensional ratio
$\lambda'=\Lambda_0/p \bar{m}$, with $\Lambda_0$ is the light
wavelength and $\bar{m}\approx 1.68$ the average refractive index,
shows the position of the bandgap, see Fig.~\ref{deVries plot}.

Importantly, in our system the initial cholesteric pitch $p_0$ was
$\sim 496$nm (labelled on the plot), so all the subsequent
action (dopant removal and the subsequent chiral intake) takes place
on the inside of the bandgap.
To find the current value of pitch $p$, which is affected by the
amount of chiral dopant in the network, we need
the approximate result derived in the earlier work,
represented by the solid line in
Fig.~\ref{deVries plot}:
\begin{equation}
p \approx -\frac{\pi \Delta m^2 + \sqrt{\pi^2
\Delta m^4+ 16 \bar{m}^2 \Lambda_0^2 (d\Psi/dz)^2}}{8 \bar{m}^2
(d\Psi/dz)} \label{pitch=f(Dn,psi)}
\end{equation}
This interpolated model will serve us for the rest of this work,
to help extracting the values of effective cholesteric pitch, as a
measure of phase chirality, from the measured $d\Psi/dz$ and the
deduced $\Delta m$.

\section{Methods}
\subsection{Preparation of imprinted elastomers}
\begin{figure}[h]
\resizebox{0.45\textwidth}{!}{\includegraphics{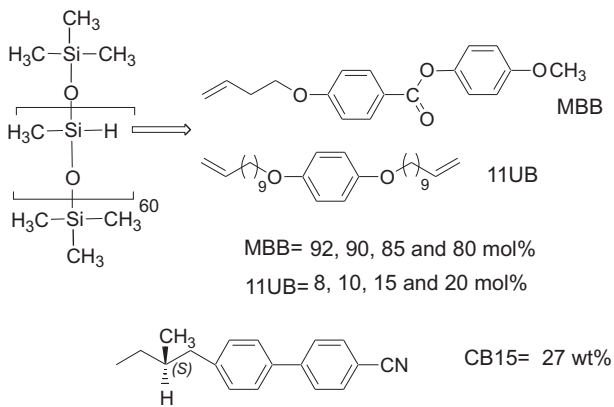}}
\caption{Chemical composition of the imprinted network. }
\label{ali}
\end{figure}

The preparation of imprinted polysiloxane side-chain cholesteric
elastomers follows the pioneering work of Kim and Finkelmann
\cite{Kim:01}, which obtains monodomain cholesteric elastomers by
uniaxial de-swelling during crosslinking. (Monodomain textures,
with the cholesteric pitch uniformly aligned along the optical
path, are essential for the study of giant optical rotation, see
below). The mesogenic group (4'-methoxyphenyl
4-(buteneoxy)benzoate, MBB) and the crosslinker
(1,4-di(11-undeceneoxy)benzene, di-11UB), both synthesized
in-house, with the molar ratios of 9.2:0.8, 9:1, 8.5:1.5 and 8:2,
doped with a fixed concentration (27\% of total weight) of chiral
compound (4-(2-methylbutyl)-4'-cyanobiphenyl, CB15), from Merck,
were reacted with polymethylsiloxane chains in toluene,
Fig.~\ref{ali}. After evaporation of the solvent and completion of
crosslinking, cholesteric elastomers were obtained -- with the
same helical pitch $\pi/q_0$, irrespective of the crosslinking
density.

In order to remove the chiral dopant CB15, the material is placed
in a large volume of non-chiral solvent (acetone) leading to a
diffusion of CB15 from the network in response to a concentration
gradient.

\subsection{Weak optical rotation}
In order to topologically imprint the helical director in the
liquid crystal elastomer, network has to be crosslinked in the
presence of chiral dopant, which then is completely removed from
the material, cf. Fig.~\ref{models}. This has been achieved by
placing the crosslinked gel, swollen with chiral dopant (in our
case 27wt\% CB15) in a large volume of nonchiral solvent (in our
case -- acetone). In response to a concentration gradient, chiral
dopant diffuses from the network to the bulk of achiral solvent,
inducing an small increase of its rotatory power
$\widetilde{\Psi}$.

\begin{figure}[h]
\resizebox{0.4\textwidth}{!}{\includegraphics{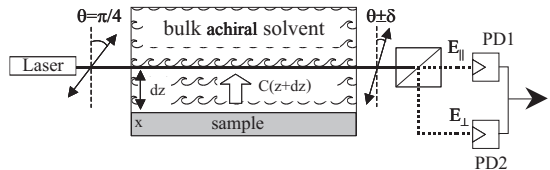}}
\caption{Differential optical apparatus to measure the small
optical rotation in bulk solvent, arising due to the diffusion of
chiral dopant from the network. The signal is proportional to the
difference between two photodiode readings, $(E_{\|}-E_{\bot})$.}
\label{set-up2}
\end{figure}

To detect very small variations of Faraday effect (optical
activity) of a chiral molecules in dilute solution, a sensitive
apparatus is needed. Our technique relies on the differential
method. An incident polarized light (laser He-Ne, Melles-Griot) is
decomposed, by using a polarizing cube beam splitters
(Melles-Griot), into its two components of electric field vector:
parallel $E_{\parallel}$, or $E_{x}$, (transmitted) and
perpendicular $E_{\perp}$, or $E_{y}$, (reflected) to the plane of
incidence. Finally, the angle of rotation $\widetilde{\Psi}$ of
the plane polarization from its incident direction is measured
with a differential detection by two photodiodes facing each
other: $\widetilde{\Psi} = \arcsin (\frac{E_{x}-E_{y}}{E\sqrt{2}
})$ where $E_{x,y}$ are the amplitudes of  signal for the two
orthogonally polarized beams (Fig.~\ref{set-up2}).

\subsection{Kinetics of dopant release}
\begin{figure}[h]
\resizebox{0.35\textwidth}{!}{\includegraphics{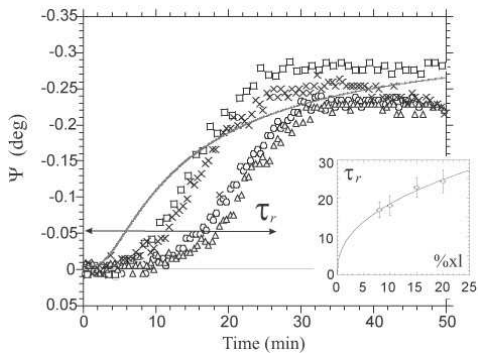}}
\caption{Release of CB15 from the network for different
crosslinker density (triangles: $20\%$;  circles: $15\%$; crosses:
$10\%$; squares: $8\%$ crosslinking density). The small angle of
rotation $\Psi$ is measured by its direct proportionality to the
differential detection $(E_x-E_y)$. Solid line shows the
corresponding Fickian diffusion dynamics. Inset: Variation of the
released time $\tau_{r}$ with crosslinker density; the line (a
square-root fit) is only a guide the the eye. } \label{release}
\end{figure}

 The
variation in space and time of the chiral dopant concentration $C$
in the bulk of the solvent, is due to the linear diffusion
(without convection) between two parallel planes $z$ (sample
surface) and $z+dz$ (beam path, $dz \sim 1$mm).
 Figure~\ref{release} shows the evolution of
$\widetilde{\Psi}$, for CB15 diffusing from networks with
different crosslink density ($8\%$, $10\%$, $15\%$ and $20\%$)
placed in an acetone bath. For all materials, the value of
$\widetilde{\Psi}$ increases and saturates at $\widetilde{\Psi}
\approx -1.5$deg.  For this ordinary isotropic Faraday effect
$\widetilde{\Psi}=\varphi d\beta$, where $\varphi$ is the solute
concentration, $d$ the optical path and $\beta$ the twisting power
of the solute. The negative absolute value of the angle
corresponds to the clockwise rotation of polarization plane, which
is the molecular property of CB15, and the phase property of the
cholesteric helix at this wavelength.

However, the kinetics of this process is more complex than the
simple Fickian diffusion, with the concentration $ \propto \exp(-z^{2}/4Dt)$.
We find the initial ``delay'' time for the distance of 1mm of
order 6-10min as in Fig.~\ref{release}. After the given delay, the
time to reach the saturation is of the same order of magnitude
$\sim$~10min for all materials. If we assume the CB15 diffusion
through a layer of acetone, after its release from the swollen
gel, is a simple diffusion, then its constant can be estimated as
$D \sim (dz)^2/\tau_r \sim 5 \cdot 10^{-5}\hbox{cm}^2/s$ (a
typical value for organic liquids of small molecules. The solid line on
the plot illustrates the Fickian law for this diffusion constant,
demonstrating the discrepancies: too long a delay and too fast
onset of saturation after that in our experimental system. Clearly the
observed kinetics is dominated by the gel de-swelling.

The characteristic times $\tau_{r}$ of reaching the solvent
saturation, obtained from Fig.~\ref{release}, are plotted against
the network cross-link density (Inset). The weak dependence of
$\tau_{r}$ on the crosslink density, and the values of time
scales, indicate the role of gel de-swelling dynamics (resulting
from the competition elasticity, diffusion and mixing, which only
recently becomes better understood \cite{doi-swell}. The small
difference in CB15 concentration in the saturated solvent between
the samples of different crosslinking density is likely to be due
to the differences in corresponding constants controlling these
effects.

\subsection{Large optical rotation}

\begin{figure} 
\resizebox{0.4\textwidth}{!}{\includegraphics{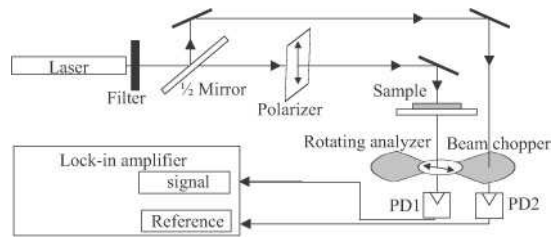}}
 \caption{
Optical set-up for measuring large optical rotation is based on
rotating analyzer to measure the optical rotation $\Psi$ from the
imprinted network on which a droplet of solvent has been deposit.
} \label{set-up1}
\end{figure}

The rotation $\Psi$ of plane-polarized light passing through a
system with helically modulated birefringence, such as the
cholesteric liquid crystal (whether natural or imprinted) can be
determined experimentally by using a dynamical method
\cite{rotref} based on measuring the phase difference between the
split parts of linearly polarized  beam, one passing through the
sample and the rotating analyzer, the other through the optical
chopper (providing the reference signal to lock on),
Fig.~\ref{set-up1}. In contrast to the differential method
described above, this technique is suitable for measuring large
rotation angles and

The laser beam (He-Ne laser, $\lambda_{laser}=633 nm$,~$30mW$,
from Melles-Griot) is split by a $1/2$-mirror. The straight part
is polarized, passes through the sample and then through the
analyzer rotating at a fixed frequency $\sim 16 Hz$, then entering
a detector. The sinusoidal signal is recorded with its phase
determined by the rotation through the sample. The other part of
the beam is sent around and through the chopper paddles mounted on
the same rotating disc. This stepwise signal provides the
reference, and the phase difference $\Delta \Theta$ between the
two beams is measured by an integer number of periods with a
lock-in amplifier (Stanford Research) and corresponds directly to
the optical rotation $\Psi$ from which the effective cholesteric
pitch $\langle p \rangle$ is then calculated. The sample is
deposited onto a glass coverslip and a solvent droplet of known
volume ($10 \mu l$) is placed on it, with the beam spot through
the middle. Finally, the conversion of the raw rotation angle
$\Psi$ into the relevant rotatory power $d\Psi/dz$ is obtained by
measuring independently the thickness variation of the network
with the solvent concentration $\phi$. This can be achieved by
simply measuring the weight of the swollen imprinted network as
function of time $t$ from its isotropic state ($t=0$) to
$t\rightarrow\infty$, since its area is conserved in the $x$-$y$
plane and would only change of thickness along $z$ as function of
$\phi$ by $\gamma_{z(\phi)}=(1+\phi)$.

The solvents used for swelling the dry imprinted network in situ
were, respectively, toluene:hexane mixture (achiral solvent, ratio
1:6, from Acros) and 2-Bromopentane (racemic mixture of right- and
left- isomers, from Acros).
%

\section{Imprinting and Stereo-selectivity}
\begin{figure}
\resizebox{0.48\textwidth}{!}{\includegraphics{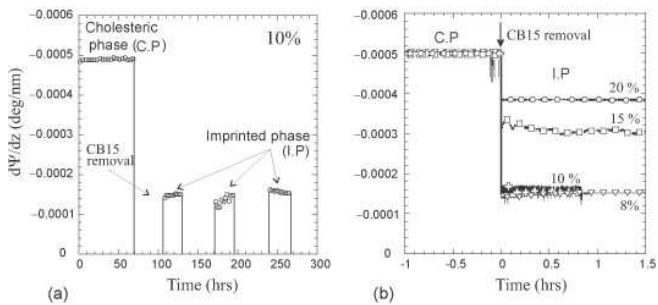}}
\caption{Residual phase chirality --imprinted phase-- after the
complete removal of the CB15 chiral dopant, represented by the
optical rotation rate $d\Psi/dz$. (a) Repeated flushes with
acetone for the $10\%$ crosslinked network show that the final
value of $d\Psi/dz$ remains stable. (b) The amount of retained
phase chirality is a function of the network crosslinked
density.} \label{imprinting proof}
\end{figure}
In this section we demonstrate the topological imprinting of the
helix in networks from which the chiral dopant (CB15) has been
completely removed. For this we detect the rotatory power
$d\Psi/dz$ of the elastomer sample (prepared such that the axis of
the imprinted helix is aligned perpendicular to the elastomer
film, in the geometry shown in Fig.~\ref{helix} and investigated
in \cite{Kim:01,Cicuta:02}). In a material with no intrinsic
molecular chirality the rotatory power is determined only by the
remaining cholesteric modulation of uniaxial dielectric constant
in an imprinted helix.
 Figure~\ref{imprinting proof}(a) shows the rotatory
power measured after repeated flushes of the elastomer (here, a
$10\%$ crosslinked sample) with an achiral solvent (acetone). In
the periods when the gel is highly swollen, it is isotropic and we
register no significant optical activity. On drying the solvent
the liquid crystalline phase returns and its remaining imprinted
helix produces a rotatory power $d\Psi/dz$, that is relatively
constant after each swelling/drying cycle (which proves that there
is no CB15 left in it after the first removal). The rotatory power
of the imprinted helix is lower than the initial value in the
natural cholesteric state in the presence of CB15 chiral dopant,
which means that in this material the critical concentration
$\phi^*$ is small but nonzero (such as for curve [{\sf a}] in
Fig.~\ref{models}).

We performed the same experiment in the networks with different
crosslink density ($8\%$,$10\%$,$15\%$
and $20\%$). The theory predicts that the remaining phase
chirality, measured by the wave number $q$ of the imprinted helix,
should increase as the parameter $\xi
q_{0}$ decreases and the nematic director is more strongly
anchored to the elastic matrix. In agreement, we find that the
amount of retained phase chirality decreases with the cross-link
density, as shown in Fig.~\ref{imprinting proof}(b).

\begin{figure}
\resizebox{0.48\textwidth}{!}{\includegraphics{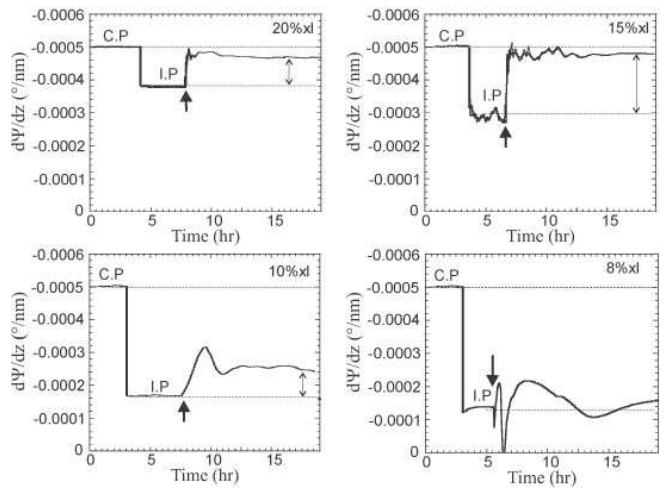}}
\caption{Variation of the rotatory power $d\Psi/dz$ for $8\%$, $10\%$,
$15\%$ and $20\%$ crosslinked networks in their imprinted state.
At time $t=0$ the racemic mixture is deposited on the dry
imprinted (I.P) samples (indicated by arrows). For both $15\%$ and
$20\%$ cross-link density
networks, the initial helical pitch (cholesteric phase, C.P) is
restored almost completely by stereo-selective absorption of chiral
molecules. For lower crosslink densities ($8\%$,
$10\%$), the variation of $d\Psi/dz$ presents a more erratic
behavior.} \label{stereo}
\end{figure}

 The resulting elastomer still retains a residual
 macroscopic phase chirality even with only centrosymmetric
 molecules left in the network. This frustrated system cannot be called a
 cholesteric liquid crystal, because this imprinted helix is not
 spontaneously formed, but is retained as a compromise between
 the untwisting trend of the nematic order and the rubber elastic
 resistance of the network to any internal deformation.

How the dry imprinted sample would behave in presence of a racemic
mixture (a $50/50$ proportion of left- and right-handed
molecules)? One naturally expects, and the first attempt on theory
\cite{Mao:01b} indeed predicts, that the elastically frustrated
network will have an ``opportunity'' to relieve its internal stress
by allowing the helix to wind more. This opportunity will be
offered to the liquid crystalline system if it preferentially absorbs
the enantiomer with the ``correct'' twisting power, matching that of CB15
(which would act as a new chiral dopant, producing a new value of $q_0$).

\begin{figure}
\resizebox{0.28\textwidth}{!}{\includegraphics{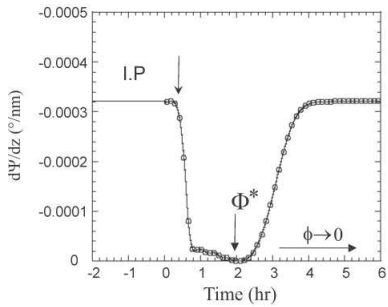}}
\caption{Variation of the rotation rate $d\Psi/dz$ for a $15\%$
cross-linked imprinted network swollen by an achiral solvent,
which is then allowed to dry. $\Phi^{*}$ is the critical solvent
concentration below which the nematic order returns to the system
(see Section~\ref{NO}).} \label{15xlachiral}
\end{figure}

From the results shown in Fig.~\ref{stereo}, we observe that the solvent
adsorption by the network affects strongly its macroscopic helical
phase, depending mainly of the strength of the polymer matrix
(cross-link density). The experiment is straightforward: the
imprinted network is swollen in the racemic solvent (at a point
in time labelled by the arrow in the plots) and then
allowed to dry again. For densely cross-linked networks ($15\%$
and $20\%$, Fig.~\ref{stereo}), the macroscopic helical phase
given by the rotatory power $d\Psi/dz$ increases and saturates to
a value close to what it was initially cross-linked in
(cholesteric phase, C.P). This is the signature of
stereo-selective swelling of imprinted networks. By selectively
retaining a sufficient amount of chiral enantiomer, which agrees with
the imprinted handedness of the network (and rejecting the
molecules with opposite handedness), imprinted networks can return
their residual helical pitch to the natural one corresponding to
the cholesteric phase (in presence of CB15 dopant). In order to
test this stereo-selective potential of imprinted networks, we did
the same experiment with a stereo-neutral solvent.
Figure~\ref{15xlachiral} clearly shows no effect on the network as
the repeated cycles in Fig.~\ref{imprinting proof}(a)   also
indicate. After the complete evaporation of the achiral solvent,
the rotatory power $d\Psi/dz$ returns back to the initial value
corresponding to the frustrated imprinted state.

For samples with weaker cross-linking, such as $8\%$ and
$10\%$, the value of $d\Psi/dz$ does not return back to the value
corresponding to the cholesteric phase, reflecting a much less
stereo-selective efficiency (Fig.~\ref{stereo}).  After
swelling both weakly imprinted networks in a racemic mixture, the phase chirality
monitored by $d\Psi/dz$ presents great instability and high
amplitude of alternating helical pitch as the network ``tries'' to
resolve its internal elastic frustration and the solvent content
(this erratic behaviour is more pronounced for the $8\%$ crosslinked
sample). These erratic oscillations are qualitatively reproducible in other
experiments and in repeated swelling cycles of the same sample.
In fact, similar oscillations are observed in many
situations of deswelling liquid crystalline polymer networks.
Although no full theoretical explanation exists, this effect may
be attributed to nonuniform distribution in space and time of
coupled solvent density and local nematic order, as we discuss in the
following section.

It appears that there is an optimal crosslinking density for imprinting,
if one aims to maximize the stereo-selective effect, around 15\% in our materials.
Figure~\ref{stereo}
suggests that at weaker linked networks the imprinting is clearly not strong
enough to produce a sufficient and reliable chirality transfer
and the resulting stereo-selectivity is small. However, in a more densely
linked 20\%xl network, although the imprinting is much more effective
(nearly all of the original C.P. helix remains in the I.P.), the swelling
capacity of it is not enough and the ``window'' of rotation rate gap between
C.P. and I.P values is small.

\section{Helicity and nematic order} \label{NO}

\begin{figure}
\resizebox{0.3\textwidth}{!}{\includegraphics{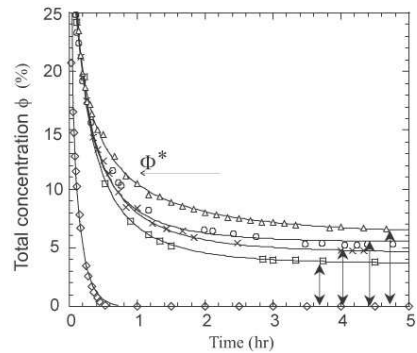}}
 \caption{Decreasing solvent content $\phi(t)$ on drying imprinted
 networks with different
crosslink density -- $8\%$xl (squares), $10\%$xl (crosses),
$15\%$xl (circles) and $20\%$xl (triangles).
For comparison, the diamond symbols show the
drying the 15\%xl network after swelling in the achiral solvent.
$\Phi^*$ labels the concentration
at which the liquid crystal order re-emerges in the system; solid
lines are the exponential fit with and without the saturation constant
at $t=\infty$ for different cases.} \label{phi-all}
\end{figure}

What is the influence of the local nematic order
$Q$ on the stereo-selective separation? As one adds a
non-mesogenic solvent to a liquid crystalline network, even a
small proportion of it reduces the local order parameter $Q$ and thus
affects the chiral order parameter $\alpha$ defined by MW.
Figure~\ref{phi-all} shows the amount of solvent in such swollen
imprinted networks, as it is allowed to dry.  The results are
shown for $\phi(t)$ in \% value to the weight of the dry imprinted
network. We specifically label the level of concentration, $\Phi^* \sim 12\%$,
at which
the highly swollen isotropic gel first returns to the liquid crystalline
state. One can tell, both visually and from the exponential fits,
how much solvent is retained by each network.
The saturation level at $t\rightarrow\infty$ is $\Delta\phi\approx 3.5\%$,
$\Delta\phi\approx 4.5\%$, $\Delta\phi\approx 5\%$ and
$\Delta\phi\approx 6\%$, for $8\%$, $10\%$, $15\%$
and $20\%$xl networks, respectively. The $15\%$xl
network swollen with a stereo-neutral solvent dries completely,
$\Delta\phi=0$.

In order to study the competition between the local nematic order
$Q$, which is being diluted by solvents and disappears altogether
above $\Phi^*$, and the macroscopic phase chirality, we need to know
how $d\Psi/dz(\phi)$ and $Q$ vary as
function of $\phi$.  We would prefer to measure the nematic
order by the value of local optical birefringence $\Delta m(\phi)$,
which are accurately in linear relation with each other, $\Delta m =
{\rm const}\cdot  Q$ \cite{deGennes93}.  We can easily obtain the
constant factor by calibrating this relation against a separate
Xray scattering image, although we will not need this specifically
in this paper.

However, it is nearly impossible
to independently measure the local birefringence $\Delta m$, or
equivalently, the local nematic order parameter $Q$, of an
elastomer imprinted with a helical texture. The difficulty is the
same as to measure $Q$ in a polydomain nematic.
As a nearest compromise, we measure $\Delta m$ on a
chemically similar aligned nematic liquid crystal elastomer (Fig.~\ref{ali},
aligned and crosslinked without CB15 dopant) and assume that
its value and variation with $\phi$ would be the same in a
cholesteric. It is not a totally unreasonable assumption: the
degree of nematic order is very reliably $Q \sim 0.5 \pm 0.1$ for
most nematic liquid crystal materials (apart from main-chain
polymers, which is not our case). The refractive indices depend
more strongly on the molecular structure, varying between, say,
$1.45$ and $1.85$ in different nematic materials.
As a confirmation of our choice, the clearing temperature of this
nematic material,  $T_c   \approx 90^{\rm o}$C, is similar to that
of the cholesteric elastomers.

\begin{figure}
\resizebox{0.37\textwidth}{!}{\includegraphics{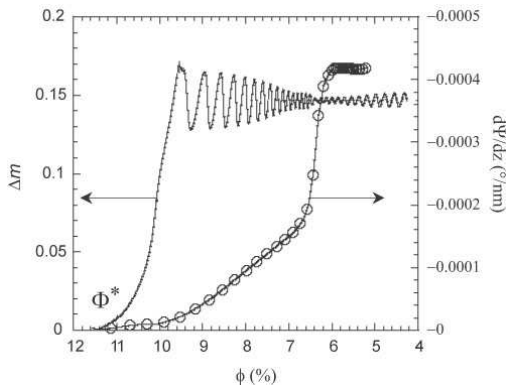}}
\caption{Superposition of birefringence $\Delta m$ (solid line)
and rotation rate $d\Psi/dz$ (circles) for a $15\%$xl imprinted
network as function of decreasing concentration $\phi$, in a
racemic environment.} \label{15xldpsi-Dm}
\end{figure}

In the following analysis, we concentrate on the
$15\%$ crosslinked network. Figure~\ref{15xldpsi-Dm} shows the
parallel results for the evolution with solvent content $\phi(t)$
of the rotation rate (a characteristic of the current helical pitch)
and the birefringence (or, equivalently, the order parameter $Q$,
obtained in the way described above).
First of all, in order to avoid any problems on
solvent concentration mapping, we brought the imprinted network to
its isotropic state above the critical solvent concentration
$\Phi^{*}$ which corresponds to the moment, when the sample first
becomes isotropic when it is gradually swollen by the solvent
($\Phi^{*}=12\%$ measured separately). Then the mapping of
the measured $d\Psi/dz(t)$ and $\Delta m (t)$ on the concentration
dependence $\phi=\phi(t)$ has been
obtained with an independent measurement
the weight of the sample, initially in isotropic phase
($\phi > \Phi^{*}$) and gradually losing the solvent
($\phi\rightarrow 0$), Fig.~\ref{phi-all}.

Although the results presented in Fig.~\ref{15xldpsi-Dm} are
too rich for us to interpret them fully, it appears that a  second
critical concentration of solvent can be defined, above which we
start loosing the underlying local order reducing the strength of
the imprinted phase chirality and the stereo-selective separation
is no longer effective. For a $15\%$xl network, this
happens at $\phi \sim 7\%$. This demonstrates the
importance of the local nematic order $Q$ (or $\Delta m$) on the
stereo-selective potential as the racemic solvent swells the network.
One can note
the unexpected oscillation of $\Delta m$ as $\phi \rightarrow 0$.
For the purpose of this paper, we do not discuss here this
phenomenon further. However we hypothesize
that these oscillations are caused by an instability of the
solvent concentration gradient
as one side of the sample is attached to a glass substrate (and
its area conserved).

\section{Reversibility and release}

\begin{figure}
 \resizebox{0.35\textwidth}{!}{\includegraphics{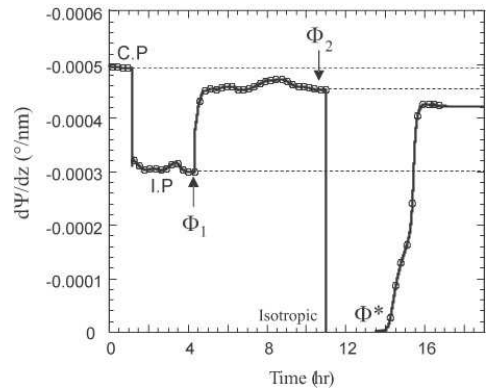}}
 \caption{Effect of high concentration solvent ($\phi>\Phi*$) on a
 $15\%$xl imprinted network: $\Phi_{1}$ and $\Phi_{2}$ correspond
 respectively to the first (small) droplet and the second (large) droplet of
 racemic solvent deposited onto the sample.} \label{15xl-solvent}
 \end{figure}

 Is the process of stereo-selectivity reproducible and reversible?
 Figure~\ref{15xl-solvent} shows the evolution with time of the
 optical rotation $\Psi$ after addition of a second droplet
 ($\Phi_{2}$) of racemic solvent. In this second exposure
 the concentration of
 solvent taken into the network exceeds the critical concentration
 $\Phi^{*}$, at which the materials becomes isotropic.
 Accordingly the optical
 rotation rapidly drops to zero;  $\Delta m=0$ and
 the material loses its coherent
 helical structure altogether. On slow evaporation, as the
 concentration of solvent decreases, the local liquid crystallinity
 returns back. Then the imprinted phase chirality is
 restored and the stereo-selective separation can be effective
 again.  As a result, some of the chiral component is retained
 again. The small difference between the resulting helical state
 in the two cases is certainly due to the difference in drying
 kinetics on the network coming back
 from the isotropic state, cf. Fig.~\ref{15xldpsi-Dm}.

\begin{figure}
 \resizebox{0.35\textwidth}{!}{\includegraphics{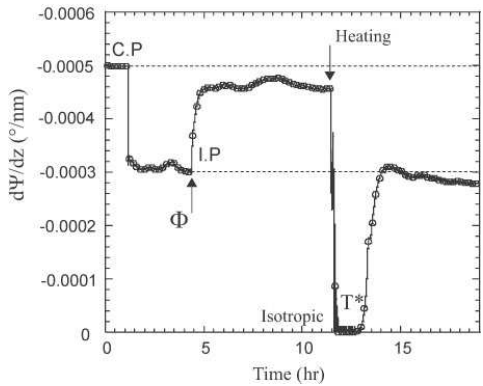}}
 \caption{Release of chiral solvent from the $15\%$xl
network by annealing. After the imprinted
 network is made to retain a portion of solvent,
 the sample is heated to above the clearing temperature,
  $T^* > 90°$C. After annealing in the isotropic phase
 for about $2$ hours, the sample is cooled to
 room temperature. The rotation returns $d\Psi/dz$
 returns back to the value corresponding to the dry imprinted state. }
 \label{15xl-temp}
 \end{figure}

 The extraction of chiral molecules trapped into the network can be
 easily achieved by heating over the critical temperature $T_c$
 of the isotropic state, as Fig.~\ref{15xl-temp}
 indicates. After a slow cooling, the optical rotation $\Psi$
 returns to the same level corresponding to the dry imprinted state
 $\langle q \rangle$, where no chiral entities are present in the
 network. Both components of the racemic mixture are able to
 evaporate with no hindrance, when the phase chirality of the
 imprinted helix is not present.
  Note that this phase chirality can be also
 mechanically tuneable if one stretches the swollen sample above
 the critical strain \cite{stretch}. This would be a more controllable way for
 repeated cycling of stereo-selective separation.


\section{Conclusions}

In this article we have reported on further details and physical
effects produced by phase chirality imprinting in elastomer
networks, as first described in \cite{Mao:01b,Courty:03}.
Although most theoretical arguments in support of the experimental
findings are based on phenomenological continuum models, the
underlying microscopic mechanism deserves some reflection. We call
this phenomenon the chirality transfer. The imprinting of a
cholesteric helix in a crosslinked network, even after removal of
chiral dopants, is unambiguous and non-controversial. What is
unusual, is that this macroscopic effect manages to influence
chemical potential acting on individual chiral molecules of the
racemic mixture. This is a universal action, quite independent on
the relative chemical structure of the original  dopant and the
subsequent racemate. A molecular picture that we require to
understand the transfer of macroscopic phase chirality down to the
molecules of the racemic mixture involves a variable-shape construction
in thermal motion. In our networks it is the strand of liquid-crystal
polymer between the crosslinks. Its average shape in the
nematic phase is a uniaxial ellipsoid, but in the presence of
cholesteric helix the statistical distribution of chain segments
becomes biased at a higher level, reflecting the additional
breaking of inversion symmetry.

It interesting that another example of chirality transfer
has recently been mentioned in the literature \cite{takezoe05}.
The new smectic liquid crystal phases (e.g. B4 and B7)
of so-called ``banana-shaped'' or bent-core
molecules are helically chiral on the submicron scale
(similar to cholesterics)
but the molecules have no molecular chirality \cite{takezoe97}.
The recent molecular model of this phenomenon involves the
flexibility of the molecules, which allows them spend
statistically more time in one of the chiral higher-energy
conformations (while the ground state shape is non-chiral).
Spectroscopic experiments appear to confirm this proposition
of the chirality transfer down to the molecular level.

A spectacular property of imprinted helical networks is their capacity to
selectively absorb and retain the component of a racemic solvent
with the matching sense of chirality. It has obvious practical implications,
however, this effect is sensitive to the
variation of the local order parameter $Q$ during the network
swelling by a solvent, since even small proportion of solvent would reduce
$Q$ and affect the stereo-selectivity. By comparing
the effects of an isotropic achiral solvent and a racemic mixture
of two opposite chiral small-molecule components, we demonstrated
how the imprinted elastomer selectively retains, after complete
restoration of the local order, the fraction $\Delta\phi$ of guest
chiral molecules in the network.  It is important to emphasize that
the chemical nature of the solvent used to test stereo-selectivity
($p$-Bromopentane) is completely different from  the chiral dopant
used originally to form the cholesteric phase (CB15).  The imprinting
of phase chirality and the subsequent chirality transfer are
universal effects, working on the mean-field level and sub-micron
length scales -- and yet clearly affecting individual solvent molecules.

One other important practical characteristic of
imprinted elastomers is that the phase chirality can be controlled
by external factors (such as temperature, solvent or
mechanical stress). Among other things, it allows the easy
triggering of release of the
chiral solvent component trapped into the network: by either
bringing the imprinted elastomer into the isotropic
phase, or by mechanically untwisting the helix.
This process, which is reproducible
and can then be reversed, makes the described stereo-selectivity
phenomenon a practical possibility for many applications (no doubt
the high surface area of a sponge of a fibrous mesh morphology would
be beneficial in any such application). On the fundamental level,
especially theoretically, much remains to be understood in the
process of chirality transfer between the length scales.

\acknowledgements
We thank Y. Mao, A. Klein, P. Cicuta and M. Warner for helpful
discussions and acknowledge financial support from EPSRC.


\end{document}